\documentclass[12pt]{iopart}

\usepackage{iopams}  
\usepackage{color}
\usepackage{graphicx}
\begin{document}

\title{Wave propagation and power deposition in blue-core helicon plasma}
\author{Lei Chang$^{1, 2}$, Juan F. Caneses$^3$, Saikat C. Thakur$^4$, and Huai-Qing Zhang$^{1*}$}
\address{$^1$State Key Laboratory of Power Transmission Equipment \& System Security and New Technology, Chongqing University, Chongqing, $400044$, China}
\address{$^2$Institute of Plasma Physics, Chinese Academy of Sciences, Hefei, $230031$, China}
\address{$^3$CompX, Del Mar, CA $92014$, USA}
\address{$^4$Department of Physics, Auburn University, Auburn, AL $36849$, USA}
\ead{zhanghuaiqing@cqu.edu.cn}

\date{\today}

\begin{abstract}
The wave propagation and power deposition in blue-core helicon plasma are computed referring to recent experiments. It is found that the radial profile of wave electric field peaks off-axis during the blue-core formation, and the location of this peak is very close to that of particle transport barrier observed in experiment; the radial profile of wave magnetic field shows multiple radial modes inside the blue-core column, which is consistent with the experimental observation of coherent high $m$ modes through Bessel function. The axial profiles of wave field indicate that, once the blue-core mode has been achieved, waves can only propagate inside the formed column with distinct phase compared to that outside. The wave energy distribution shows a clear and sharp boundary at the edge of blue-core column, besides which periodic structures are observed and the axial periodicity inside is nearly twice that outside. The dispersion relation inside the blue-core column exhibits multiple modes, a feature of resonant cavity that selects different modes during frequency variation, while the dispersion relation outside gives constant wave number with changed frequency. The power deposition appears to be off-axis in the radial direction and periodic in the axial direction, and mostly inside the blue-core column. Analyses based on step-like function theory and introduced blue-core constant provide consistent results. The equivalence of blue-core column to optical fiber for electromagnetic communication is also explored and inspires a novel application of helicon plasma, which may be one of the most interesting findings of present work. 
\end{abstract}



\maketitle

\section{Introduction}
The underlying physics of helicon discharge, which can produce high-density plasma with remarkable ionisation rate, have been attracting great research interest\cite{Boswell:1970aa, Boswell:1997aa, Chen:1997aa, Chen:2015aa, Shinohara:2018aa, Takahashi:2019aa}. Besides further measurements of TG (Trivelpiece–Gould) mode and energetic electrons, the blue-core phenomenon that plasma shrinks onto axis and emits bright light in blue colour (argon) for high input power and confining magnetic field remains mysterious and most challenging of the field\cite{Guo:1999aa, Blackwell:2012aa, Thakur:2015aa, Zhang:2021aa, Wang:2021aa, Chang:2022aa}. It represents a general bright-core mode of helicon discharge for other gases as well, for example nitrogen and helium\cite{Zhao:2017aa, Huang:2020aa}, although in different colours, and is the highest level for which the gas is fully ionised (further increased power does not enhance the plasma density but the ionisation state due to ion pumping effects\cite{Boswell:1984ab}). Four common features have been drawn from this mode: radial electrostatic confinement, light emission in phase axially, azimuthal instabilities driven by radial pressure gradients, and high-beta (beta is the ratio of particle pressure to magnetic field pressure) effects\cite{Boswell:2021aa}. This work is devoted to studying these features in terms of wave propagation and power deposition. A well-benchmarked electromagnetic solver (EMS)\cite{Chen:2006aa} will be employed for computations, referring to the recent experiments on CSDX (controlled shear decorrelation experiment)\cite{Thakur:2015aa, Thakur:2014aa}, and the step-like function theory\cite{Breizman:2000aa} will be utilised for physics analysis, followed by equivalent analogy of blue-core plasma column to optical fiber for electromagnetic communications. We shall present that the spatial features of blue-core mode observed in experiments could be well seen from the wave propagation and power deposition computations, which offer more details. Further exploration on the high-beta effects, maybe in pulsed mode\cite{Chang:2022aa, Corr:2007aa}, will be given in a follow-up work. The equivalence to optical fiber may promote novel applications of blue-core helicon plasma, an exciting and valuable finding of present work. 

\section{Numerical Scheme}
\subsection{Electromagnetic solver}
The EMS, which has been used successfully to model various plasma sources\cite{Zhang:2008aa, Lee:2011aa, Chang:2012aa, Chang:2013aa, Chang:2018aa, Chang:2020aa}, is based on two Maxwell's equations: Faraday's law and Ampere's law\cite{Chen:2006aa},
\small
\begin{equation}
\nabla\times\mathbf{E}=-\frac{\partial\mathbf{B}}{\partial t},
\label{eq1}
\end{equation}
\begin{equation}
\nabla\times\mathbf{B}=\mu_0\left(\mathbf{j}_a+\frac{\partial\mathbf{D}}{\partial t}\right),
\label{eq2}
\end{equation}
\normalsize
with $\mathbf{E}$ and $\mathbf{B}$ the wave electric and magnetic fields, respectively. The symbols of $\mu_0$ and $t$ are standard permeability of vacuum and time. The system is driven by the current density $\mathbf{j}_a$ of external antenna. Perturbations vary in form of $\exp[i(kz+m\theta-\omega t)]$, with $k$ the axial wave number, $m$ the azimuthal mode number and $\omega$ the driving frequency, for a right-hand cylindrical coordinate system $(r;\theta;z)$. The displacement vector $\mathbf{D}$ is linked to $\mathbf{E}$ via a cold-plasma dielectric tensor\cite{Ginzburg:1970aa},
\small
\begin{equation}
\mathbf{D}=\varepsilon_0[\varepsilon\mathbf{E}+i g(\mathbf{E}\times\mathbf{b})+(\eta-\varepsilon)(\mathbf{E}\cdot\mathbf{b})\mathbf{b}]. 
\label{eq3}
\end{equation}
\normalsize
Here, $\varepsilon_0$ is the permittivity of vacuum and $\mathbf{b}$ is the unit vector of external magnetic field ($\mathbf{b}=\mathbf{B_0}/B_0$). The dielectric tensor comprises three components: 
\small
\begin{equation}
\varepsilon=1-\sum_\alpha\frac{\omega+i\nu_\alpha}{\omega}\frac{\omega_{p\alpha}^2}{(\omega+i\nu_\alpha)^2-\omega_{c\alpha}^2},
\label{eq4}
\end{equation}
\begin{equation}
g=-\sum_\alpha\frac{\omega_{c\alpha}}{\omega}\frac{\omega_{p\alpha}^2}{(\omega+i\nu_\alpha)^2-\omega_{c\alpha}^2},
\label{eq5}
\end{equation}
\begin{equation}
\eta=1-\sum_\alpha\frac{\omega_{p\alpha}^2}{\omega(\omega+i\nu_\alpha)}.
\label{eq6}
\end{equation}
\normalsize
The subscript $\alpha$ labels the species of particles, i. e. ion and electron, and the plasma frequency $\omega_{p\alpha}=\sqrt{n_\alpha q_\alpha^2/\varepsilon_0 m_\alpha}$ and cyclotron frequency $\omega_{c\alpha}=q_\alpha B_0/m_\alpha$ are standard definitions. The phenomenological collision frequency $\nu_\alpha$ accounts for collisions between electrons, ions and neutrals, where background pressure is implemented. For the half-turn helical antenna considered below, $\mathbf{j}_a$ has three components: 
\small
\begin{equation}
\begin{array}{ll}
j_{ar}=0, 
\label{eq7}
\end{array}
\end{equation}
\begin{equation}
\begin{array}{ll}
j_{a\theta}=&I_a\frac{e^{im\pi}-1}{2}\delta(r-R_a)\left\{\frac{i}{m\pi}\left[\delta(z-z_a)+\delta(z-z_a-L_a)\right]\right.\\
\\
&\left.+\frac{H(z-z_a)H(z_a+L_a-z)}{L_a}e^{-im\pi[1-(z-z_a)/L_a]}\right\},
\label{eq8}
\end{array}
\end{equation}
\begin{equation}
\begin{array}{ll}
j_{az}=I_a\frac{e^{-im\pi[1-(z-z_a)/L_a]}}{\pi R_a}\frac{1-e^{im\pi}}{2}\delta(r-R_a)\times H(z-z_a)H(z_a+L_a-z). 
\label{eq9}
\end{array}
\end{equation}
\normalsize
Here, the subscript $a$ denotes the antenna, i. e. $L_a$ the length, $R_a$ the radius, $z_a$ the distance to left endplate, $I_a$ the magnitude of antenna current, and $H$ is the Heaviside step function. The boundary conditions enclosing the model are formed by assuming that the tangential components of $\mathbf{E}$ vanish on the surface of chamber walls:
\small
\begin{equation}
E_\theta(R, z)=E_z(R, z)=0, 
\label{eq10}
\end{equation}
\begin{equation}
E_r(r, 0)=E_\theta(r, 0)=0, 
\label{eq11}
\end{equation}
\begin{equation}
E_r(r, L)=E_\theta(r, L)=0,
\label{eq12}
\end{equation}
\normalsize
where $R$ and $L$ are the radius and length of chamber, respectively. 

\subsection{Domain and conditions}
\begin{figure}[ht]
\begin{center}
\includegraphics[width=0.95\textwidth,angle=0]{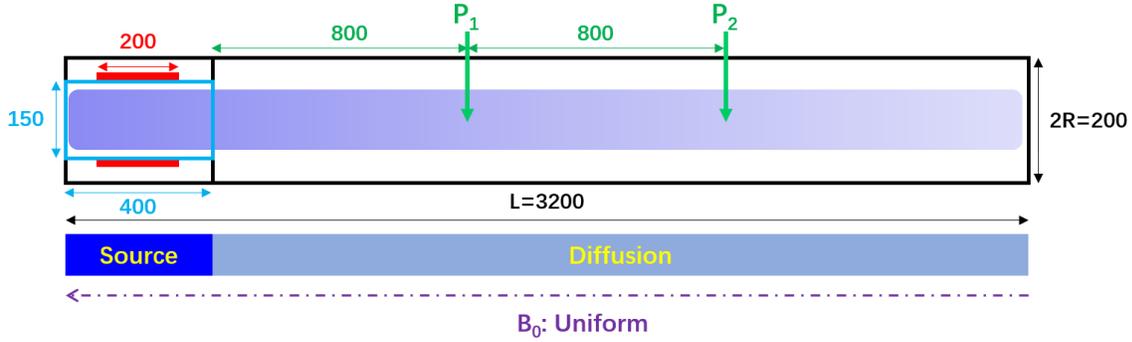}
\end{center}
\caption{Schematic of computational domain referring to CSDX\cite{Thakur:2015aa, Thakur:2014aa, Burin:2005aa}.}
\label{fg_schematic}
\end{figure}

\begin{figure}[ht]
\begin{center}
\includegraphics[width=0.49\textwidth,angle=0]{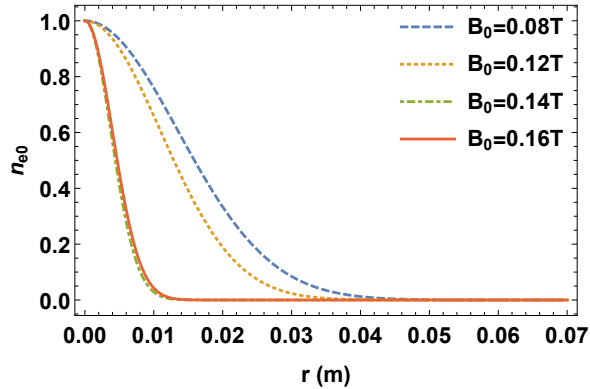}
\end{center}
\caption{Normalised radial profiles of plasma density for four external magnetic field strengths, fitted from experimental data on the CSDX\cite{Thakur:2015aa, Cui:2016aa}.}
\label{fg_gaussian}
\end{figure}

The EMS model is solved by finite difference method based on four staggered rectangular grids\cite{Chen:2006aa}. Figure~\ref{fg_schematic} shows the computational domain. It is drawn as close as possible to the recent layout of CSDX experiments\cite{Thakur:2015aa, Thakur:2014aa}, while taking into account the real operation of EMS. The source chamber (pyrex glass) in length of $0.4$~m and diameter of $0.15$~m connects the diffusion chamber (stainless steel) of length $2.8$~m and diameter $0.2$~m coaxially. Two diagnostic ports are located at $z=1.2$~m ($P_1$) and $z=2$~m ($P_2$), respectively, in the diffusion chamber to measure the cross-sectional parameters. 
The source plasma is generated by a half-turn helical antenna in length of $0.2$~m and diameter of $0.15$~m, for which the current density is expressed by Eq.~(\ref{eq7})$\sim$Eq.~(\ref{eq9}). The antenna is driven by a radio-frequency (RF) power supply with fixed frequency of $13.56$~MHz and maximum power of $5$~kW. Here, we also fix the current magnitude to $12$~A throughout the paper for accurate comparison, which is within the power capacity. The filling gas is argon and the external magnetic field is uniform with typical strengths of $0.08$~T, $0.12$~T, $0.14$~T and $0.16$~T. Accordingly, four normalised radial profiles of plasma density are constructed, as shown in Fig.~\ref{fg_gaussian}. They are fitted from experimental data\cite{Thakur:2015aa, Cui:2016aa} with expressions: $\exp[-2746.46 r^2]$ for $0.08$~T, $\exp[-4167.88 r^2]$ for $0.12$~T, $\exp[-35526.8 r^2]$ for $0.14$~T, and $\exp[-31493.9 r^2]$ for $0.16$~T. We can see that the plasma shrinks onto axis with increased magnetic field, i. e. larger gradient in radius, whereas field strength higher than $0.14$~T yields little difference. Other conditions are also set to be the same to the CSDX experiments\cite{Thakur:2015aa, Thakur:2014aa}, including electron temperature of $4$~eV.

\section{Computed Results}
\subsection{Wave propagation}
\begin{figure}[ht]
\begin{center}$
\begin{array}{ll}
(a)&(b)\\
\includegraphics[width=0.465\textwidth,angle=0]{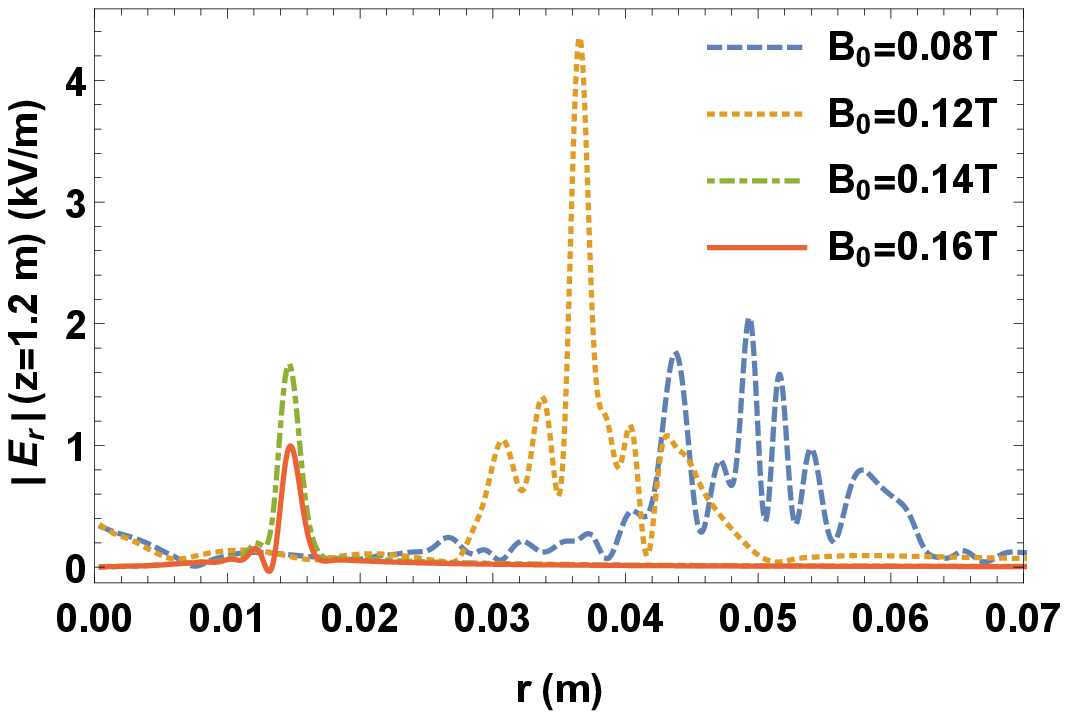}&\includegraphics[width=0.49\textwidth,angle=0]{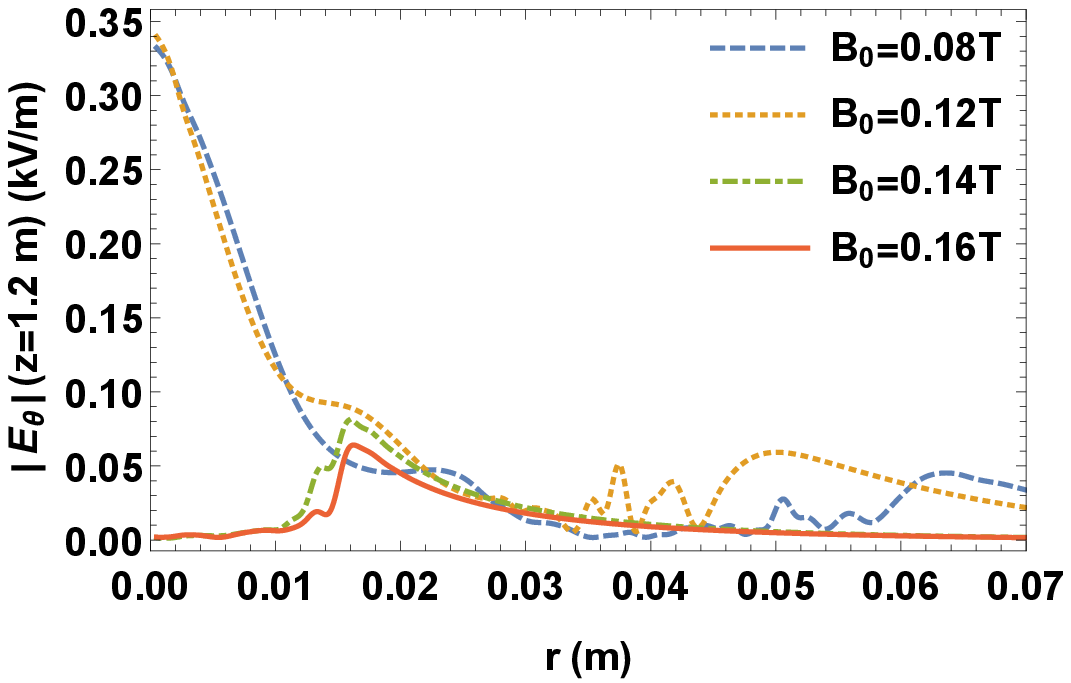}
\end{array}$
\end{center}
\caption{Radial profiles of wave electric field (a, radial component; b, azimuthal component) for four external magnetic field strengths, measured at $z=1.2$~m ($P_1$).}
\label{fg_efd_gsr}
\end{figure}

\begin{figure}[ht]
\begin{center}$
\begin{array}{ll}
(a)&(b)\\
\includegraphics[width=0.478\textwidth,angle=0]{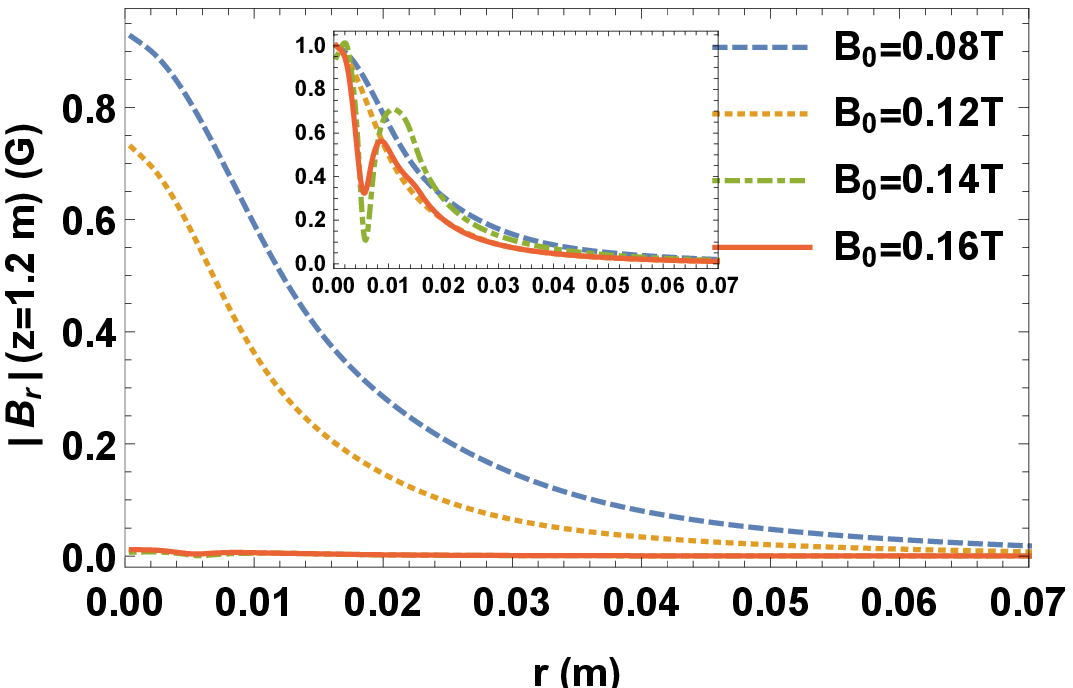}&\includegraphics[width=0.48\textwidth,angle=0]{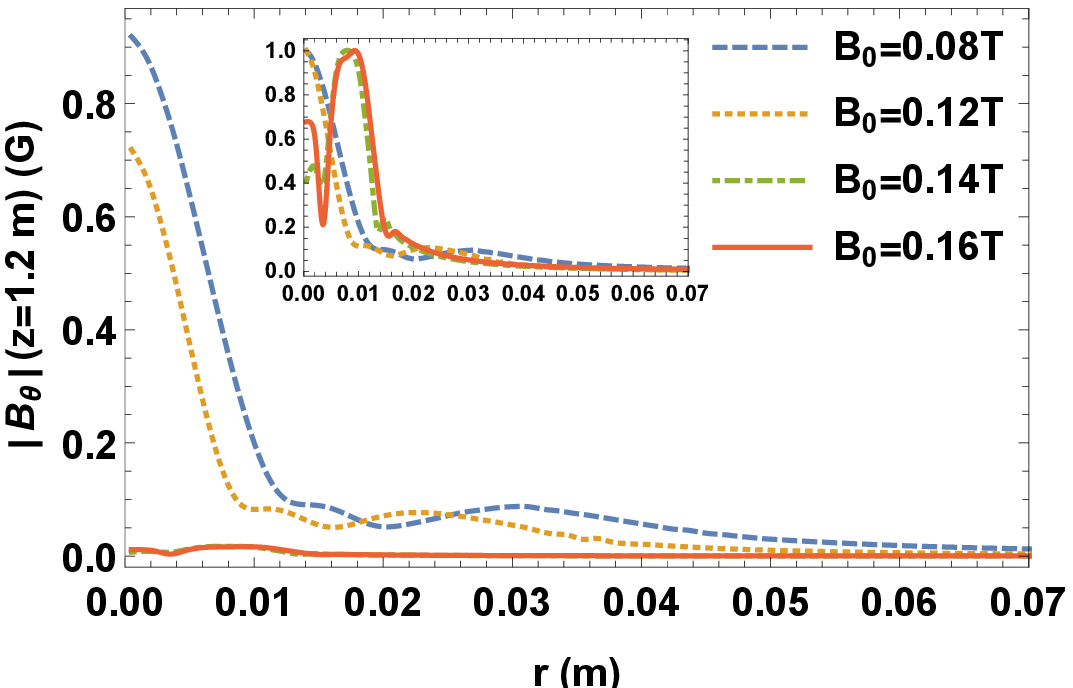}
\end{array}$
\end{center}
\caption{Radial profiles of wave magnetic field (a, radial component; b, azimuthal component) for four external magnetic field strengths, measured at $z=1.2$~m ($P_1$).}
\label{fg_mfd_gsr}
\end{figure}

\begin{figure}[ht]
\begin{center}$
\begin{array}{ll}
(a)\\
\includegraphics[width=0.75\textwidth,angle=0]{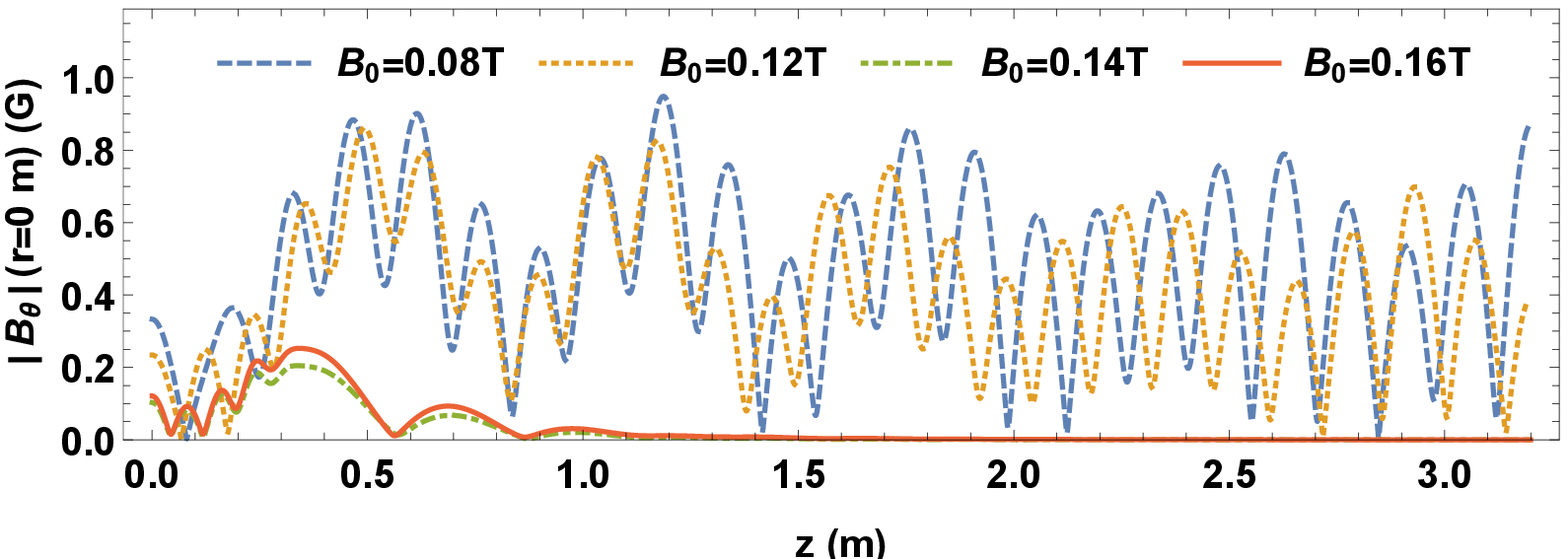}\\
(b)\\
\hspace{-0.15cm}\includegraphics[width=0.76\textwidth,angle=0]{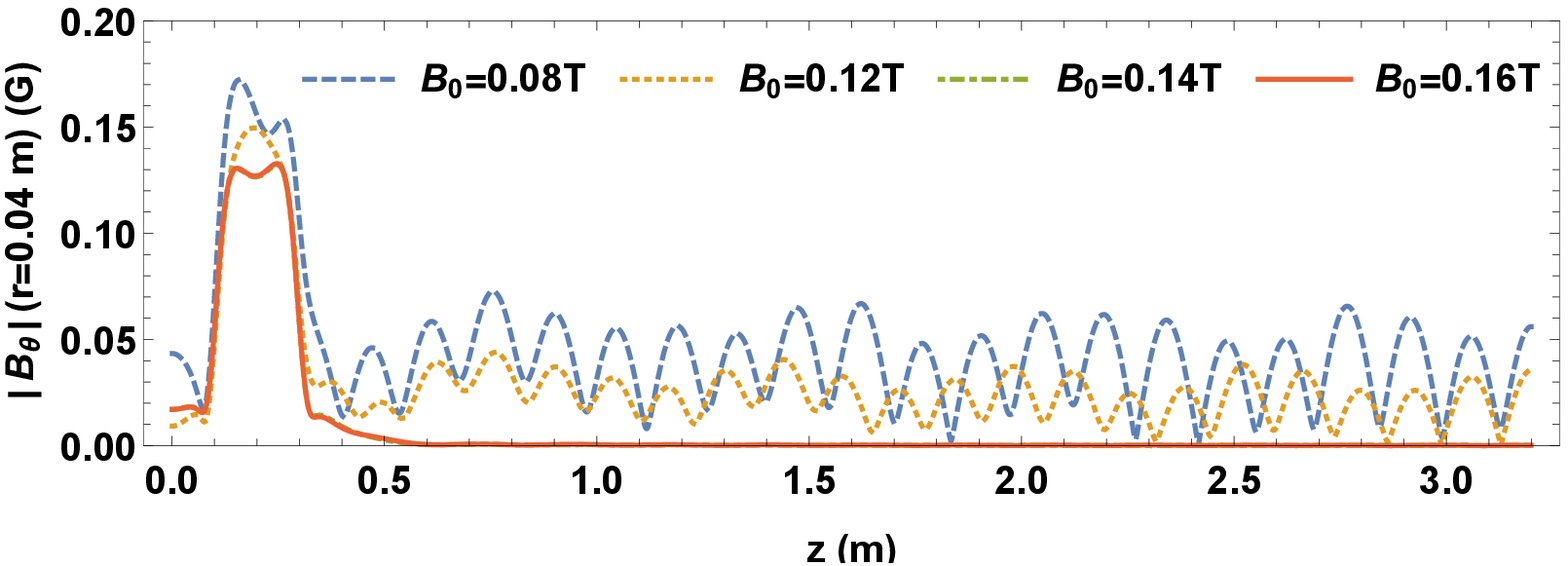}
\end{array}$
\end{center}
\caption{Axial profiles of wave magnetic field for various confining field strengths, measured inside (a, $r=0$~m) and outside (b, $r=0.04$~m) blue-core column.}
\label{fg_mfd_gsz}
\end{figure}

\begin{figure}[ht]
\begin{center}$
\begin{array}{ll}
(a)\\
\includegraphics[width=0.745\textwidth,angle=0]{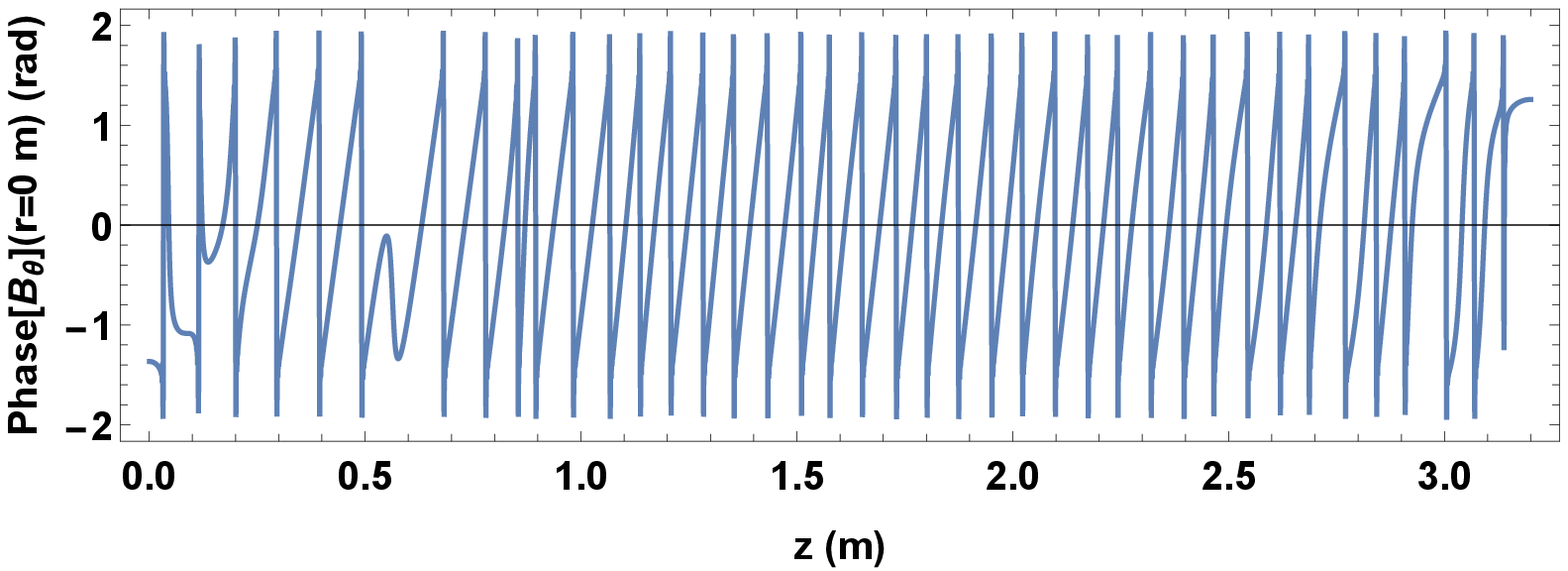}\\
(b)\\
\hspace{-0.23cm}\includegraphics[width=0.76\textwidth,angle=0]{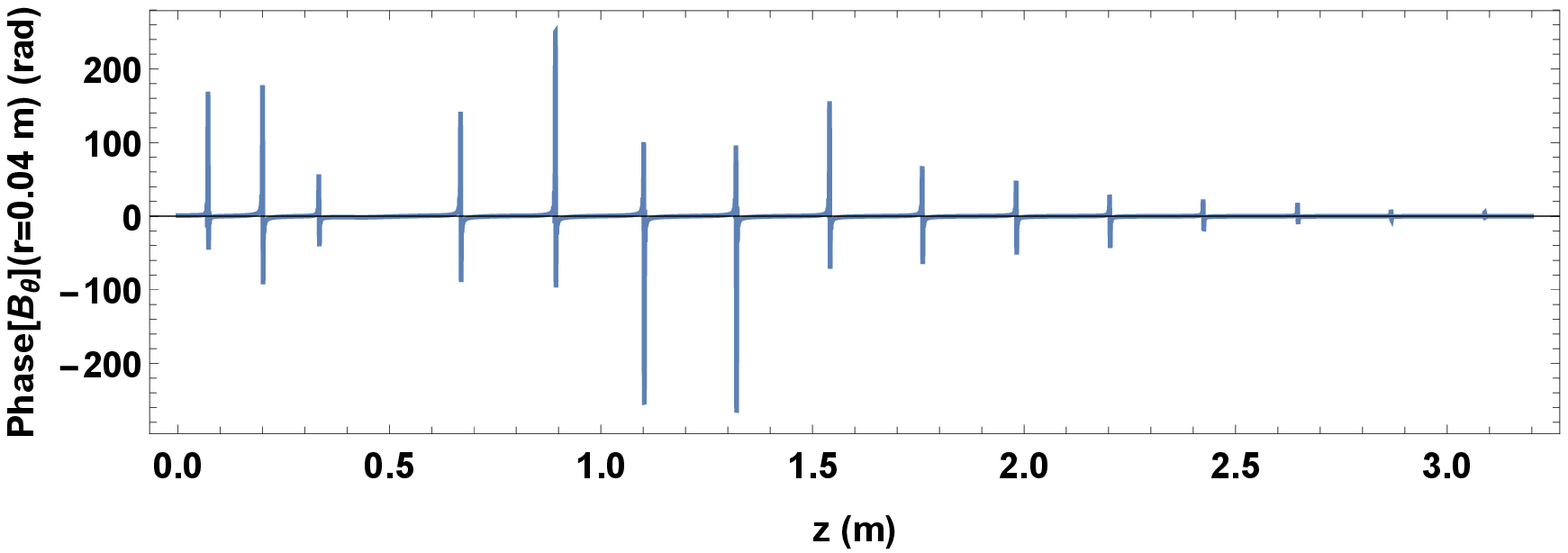}
\end{array}$
\end{center}
\caption{Phase for wave magnetic field inside (a, $r=0$~m) and outside (b, $r=0.04$~m) blue-core column ($B_0=0.16$~T).}
\label{fg_phase}
\end{figure}

\begin{figure}[ht]
\begin{center}
\includegraphics[width=0.85\textwidth,angle=0]{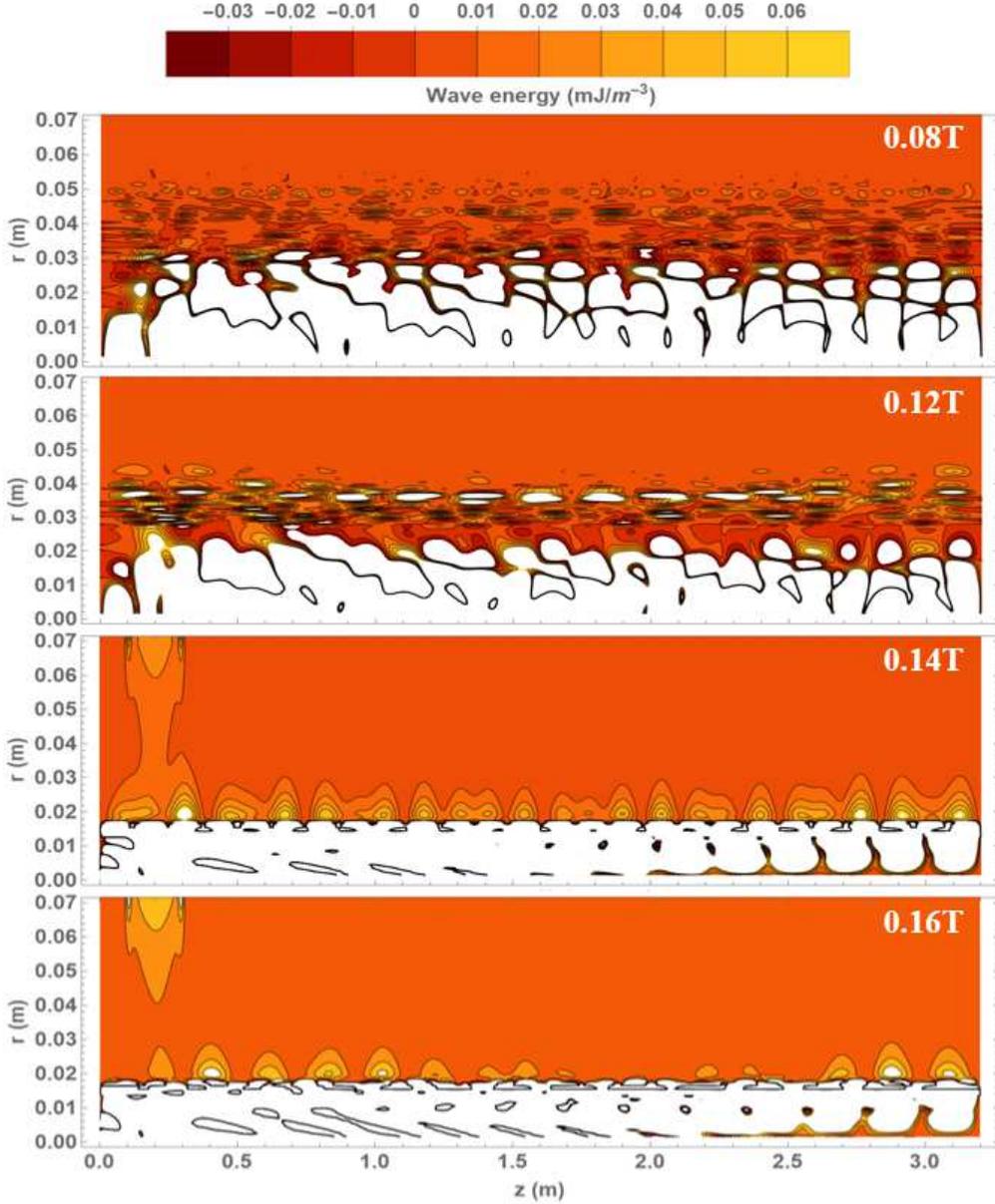}
\end{center}
\caption{Wave energy distributions for various confining magnetic field strengths.}
\label{fg_energy}
\end{figure}

\begin{figure}[ht]
\begin{center}
\includegraphics[width=0.499\textwidth,angle=0]{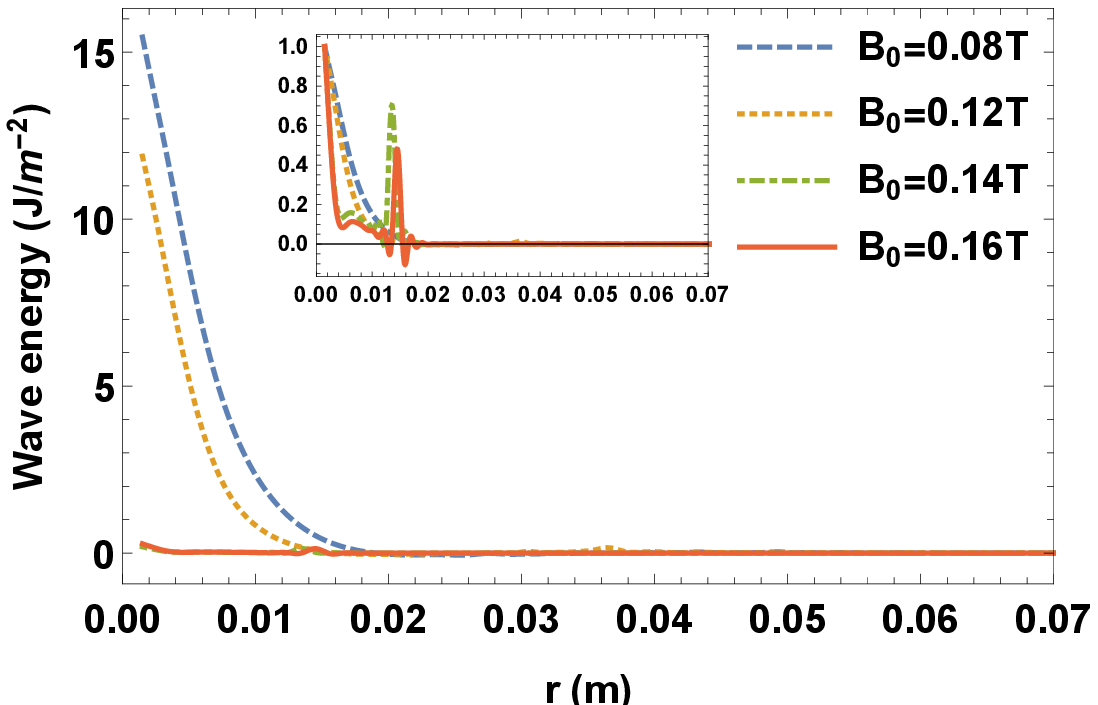}
\end{center}
\caption{Integrated radial profiles of wave energy for the four external magnetic field strengths. The inset shows normalised results.}
\label{fg_energy_gsr}
\end{figure}

\begin{figure}[ht]
\begin{center}$
\begin{array}{ll}
(a)&(b)\\
\includegraphics[width=0.46\textwidth,angle=0]{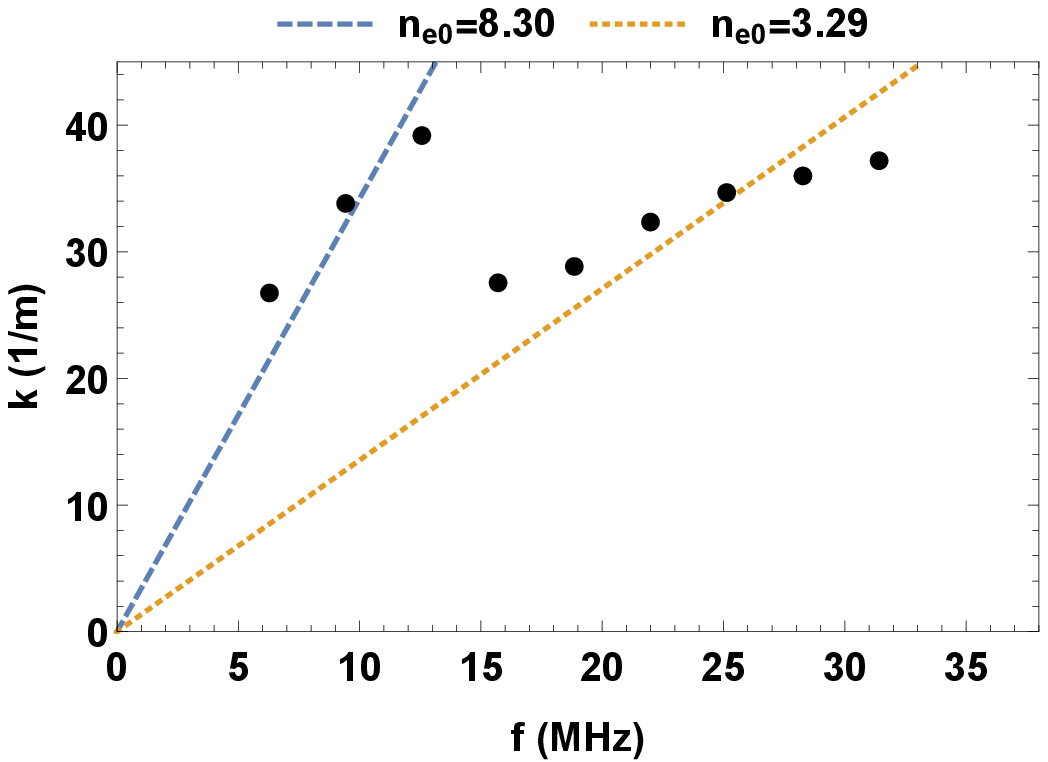}&\includegraphics[width=0.46\textwidth,angle=0]{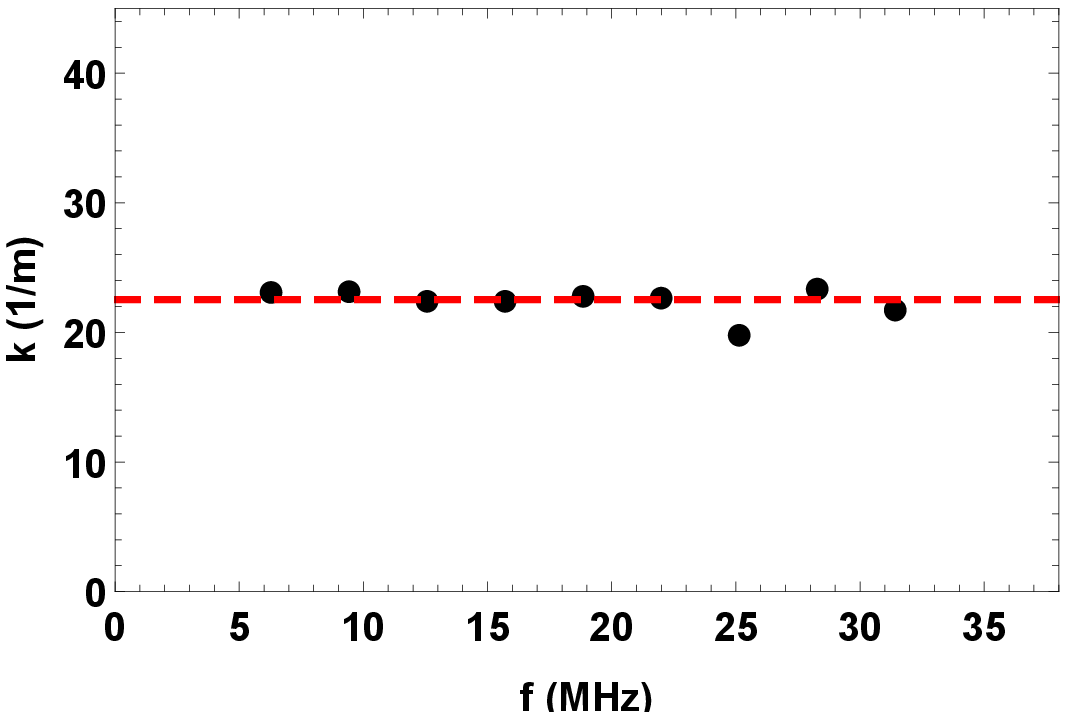}
\end{array}$
\end{center}
\caption{Dispersion relations inside (a, $r=0$~m) and outside (b, $r=0.04$~m) blue-core column for $B_0=0.16$~T (Gaussian density profile).}
\label{fg_dispersion_gs}
\end{figure}

We first investigate the wave propagation characteristics for the plasma density profiles constructed above. Figure~\ref{fg_efd_gsr} and Fig.~\ref{fg_mfd_gsr} give the radial profiles of wave electric field and magnetic field, respectively, calculated at the first port ($P_1$ in Fig.~\ref{fg_schematic}), which provides convenience for experimental verification. It can be seen that the wave electric field clearly forms a local peak around $r\approx 0.015$~m for high field strengths. Considering that this peak expels ions away and attracts electrons together on both sides, we thereby claim that it forms a local transport barrier and is an evidence of radial electrostatic confinement. Further, this temporally oscillating peak can induce ponderomotive force which may also contribute to the formation of transport barrier\cite{Lundin:2006aa, Khazanov:2000aa, Khazanov:2013aa}. Indeed, the location of this peak is very close to the experimental observation of transport barrier at $r\approx0.02$~m during the blue-core formation\cite{Thakur:2015aa, Thakur:2014aa}.
Moreover, the wave magnetic field shows additional radial modes inside $r\approx 0.015$~m for high field strengths. This is consistent with the experimental observation that very coherent high-$m$ fluctuations occur inside the blue-core column. Due to the limitation of EMS that only a single $m$ can be considered for each run of simulation ($m=1$ throughout the paper for the half-turn helical antenna employed here), however, we cannot see directly the appearance of high-$m$ structure but indirectly from the formation of multiple radial modes.
This can be inferred from the resonant eigenmodes\cite{Chang:2014aa}:
\small
\begin{equation}
B_{r}=\frac{i U}{2Q}\left[(\beta+k)J_{m-1}(Q r)+(\beta-k)J_{m+1}(Q r)\right],
\label{eq13}
\end{equation}
\begin{equation}
B_{\theta}=-\frac{U}{2Q}\left[(\beta+k)J_{m-1}(Q r)-(\beta-k)J_{m+1}(Q r)\right],
\label{eq14}
\end{equation}
\normalsize
with $U$ the amplitude constant and $Q^2=\beta^2-k^2$. It is the Bessel function $J_m(r)$ that correlates the radial and azimuthal modes, while in physics they are essentially determined by the spatial eigenmode resonance. Overall, the radial profiles of wave electric field and magnetic field show significant difference when the confining magnetic field increases from low magnitude (before blue-core formation) to high magnitude (after blue-core formation). These different propagation features before and after the blue-core formation can be also observed in the axial direction. We compute the wave magnetic field, which is easily measurable by B-dot probe in experiment, at two radial locations, namely $r=0$~m (inside blue-core column) and $r=0.04$~m (outside blue-core column). As shown by Fig.~\ref{fg_mfd_gsz}, the wave propagation is more evanescent for higher field strength with blue-core formation, and this occurs for both radial locations. Moreover, comparing the wave field profiles at these two locations, we can see that the wave propagation on axis surpasses that near edge, especially after the blue-core formation. 
To show more details, we compute the phase of wave magnetic field inside and outside the blue-core column ($B_0=0.16$~T), via $\theta=\textrm{arctan}~\textrm{Im}[B_\theta]/\textrm{Re}[B_\theta]$. Figure~\ref{fg_phase} presents the results. One can see that the wave propagation inside has identical phase in the axial direction ($z\approx 1.1\sim 2.7$~m away from antenna and endplates), which is consistent with the light emission in phase axially observed in experiment\cite{Boswell:2021aa}, whereas there seems no regular phase outside but periodic singularities. 
The axial profiles of wave electric field show similar features. This implies that the existence of blue-core plasma column in a sense confines the wave propagation inside, which is illustrated more clearly by the two-dimensional contour plots of wave energy distribution shown in Fig.~\ref{fg_energy}. We can see that, with the field strength increased, the wave energy shrinks towards axis and forms a sharp boundary around $r\approx 0.018$~m, close to the location of transport barrier at $r\approx 0.02$~m observed in experiment\cite{Thakur:2015aa, Thakur:2014aa}. Interestingly, quasi-periodic structures are formed axially on both sides of the boundary layer, and the periodic length inside is close to twice that outside. This structure looks similar to the beat pattern of helicon radiation observed earlier\cite{Chang:2012aa, Caneses:2016aa}, and is consistent with the whistler bouncing at sharp plasma edge\cite{Shamrai:1996aa, Caneses:2017aa}. Moreover, the wave energy distribution becomes off-axis when it expands from the source region to the diffusion region.
Actually, for these high-field cases, i. e. $B_0=0.14$~T and $B_0=0.16$~T, there always exists an off-axis peak in the radial profile of wave energy density, as shown in Fig.~\ref{fg_energy_gsr}. The radial locations of these off-axis peaks ($r\approx 0.015$~m) are very close again to that of experimental transport barrier ($r\approx 0.02$~m)\cite{Thakur:2015aa, Thakur:2014aa}. Here, integration has been performed along the axial direction, same to the method of experimental measurement, showing the accumulated wave energy in the cross section.
To reveal the intrinsic physics of different wave propagation features inside and outside the blue-core column, we focus on the case of $B_0=0.16$~T and run the EMS code for frequency range of $f=\pi\sim 10\pi$~MHz. Utilising a Fourier decomposition method to extract the dominant $k$ from the axial profiles of wave field, we obtain the dispersion relations shown in Fig.~\ref{fg_dispersion_gs}.
For the location inside blue-core column, analytical dispersion relation\cite{Chen:1996ab} 
\small
\begin{equation}
k=\frac{2\pi R_p}{3.83}\frac{e\mu_0 n_{e}}{B_0}f
\label{eq15}
\end{equation}
\normalsize
has been also plotted with $R_p=0.02$~m (plasma radius) and $n_e=n_{e0}\times 10^{19}~\textrm{m}^{-3}$. Here, the number $3.83$ is the first non-zero Bessel root of $J_1(r)=0$, representing the first radial mode. The straight lines are fitted results. It seems that the formed blue-core column behaves as a resonant cavity, which could shift and select different modes for different ranges of driving frequency. For the location outside blue-core column, the computed dispersion relation is not linear, thus cannot be fitted with Eq.~(\ref{eq15}). Instead, we have drawn a horizontal line to label the average value of $k=22.37~\textrm{m}^{-1}$. The corresponding wave length is $\lambda=0.28$~m, close to the length of antenna ($0.2$~m), and independent of frequency, whereas the wave length inside blue-core column is much shorter and varies with frequency. This demonstrates that the wave modes inside and outside blue-core column are essentially different. As shown in Fig.~\ref{fg_mfd_gsz} and Fig.~\ref{fg_phase}, once the blue-core column has been formed, wave propagation is mostly confined inside, similar to the light propagation in optical fiber\cite{Kao:1966aa, Maurer:1973aa}. This feature is also consistent with the experimental observation of light emission in phase axially\cite{Boswell:2021aa}, and may inspire novel applications of blue-core helicon plasma, which will be analysed further in next section. 

\subsection{Power deposition}
\begin{figure}[ht]
\begin{center}
\includegraphics[width=0.49\textwidth,angle=0]{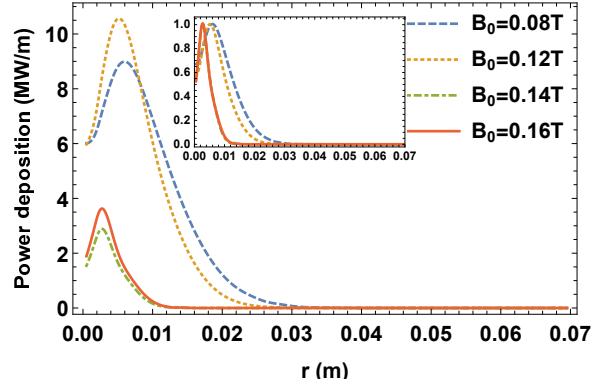}
\end{center}
\caption{Integrated radial profiles of power deposition for the four external magnetic field strengths. The inset shows normalised results.}
\label{fg_power_gsr}
\end{figure}

\begin{figure}[ht]
\begin{center}
\includegraphics[width=0.85\textwidth,angle=0]{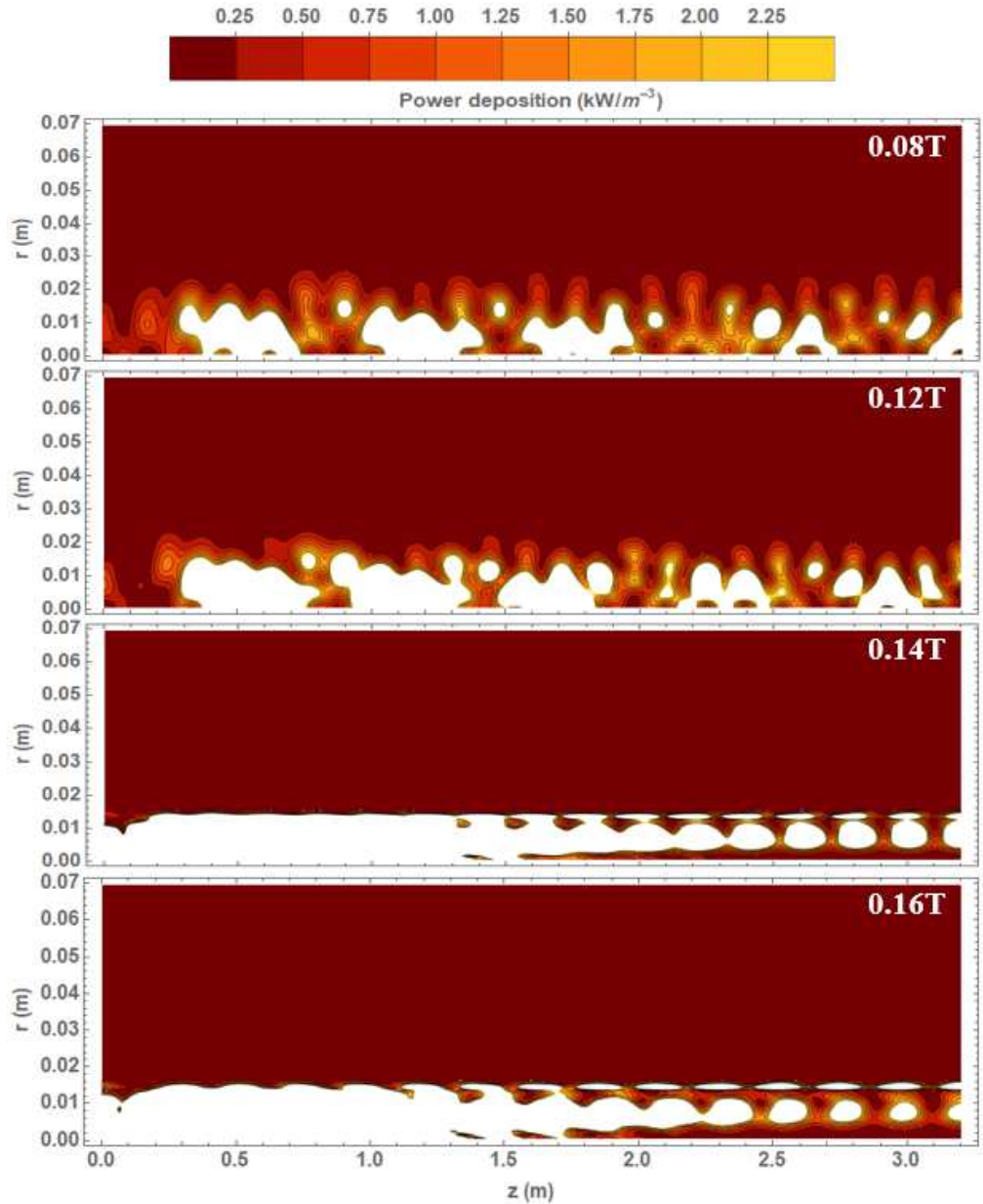}
\end{center}
\caption{Power depositions for various confining magnetic field strengths.}
\label{fg_power}
\end{figure}

Next, we explore how the power is deposited from antenna to plasma, especially when the blue-core column has been formed. Figure~\ref{fg_power_gsr} shows the radial profiles of computed power deposition, which have been integrated over the axial direction in a similar way as for Fig.~\ref{fg_energy_gsr}. One can see that the magnitude of power deposition reduces significantly when the field strength increases from $0.12$~T to $0.14$~T, indicating an evolution into distinct mode. This is consistent with the experimental observation that the discharge transits from blue-colour mode to blue-core mode\cite{Thakur:2015aa}. The decreased power for shrunk density profiles implies that certain magnitude of plasma density near edge is beneficial for power coupling, which also agrees with previous studies\cite{Chang:2016aa, Chang:2018ab, Wang:2019aa, Isayama:2019aa, Isayama:2019ab}. More interestingly, we find that the power deposition is hollow in radius for all field strengths and its peak moves closer to axis when the field strength increases. 
A full picture of power deposition is given by Fig.~\ref{fg_power}. We can see that, similar to Fig.~\ref{fg_energy}, the distribution shrinks radially towards axis when the field strength is increased and forms a sharp boundary after the blue-core establishment. The radial location of this layer ($r=0.015$~m) is slightly closer to axis than that formed by the wave energy distribution ($r=0.018$~m), but both close to the radius of measured transport barrier ($r=0.02$~m)\cite{Thakur:2015aa, Thakur:2014aa}. Moreover, the axial periodic structures are observed again inside and ouside the boundary layer; different from Fig.~\ref{fg_energy}, however, the periodic lengths are equivalent. Further, the maximum power deposition is off-axis, especially far away from the antenna in the diffusion region. 

\section{Theoretical Analysis}
\subsection{Step-like function theory}
\begin{figure}[ht]
\begin{center}
\includegraphics[width=0.4\textwidth,angle=0]{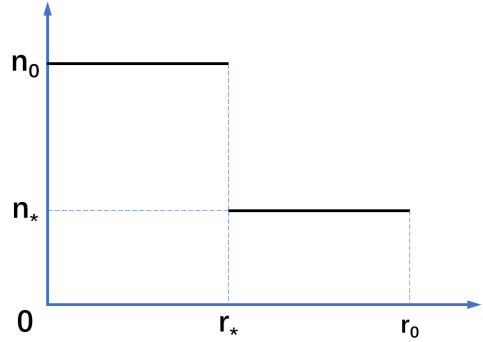}
\end{center}
\caption{Step-like density profile in radius, a schematic illustrating the concept of blue-core constant below ($C_b=C_n\times C_r=n_\ast/n_0\times r_\ast/r_0$).}
\label{fg_steplike}
\end{figure}

\begin{figure}[ht]
\begin{center}
\includegraphics[width=0.49\textwidth,angle=0]{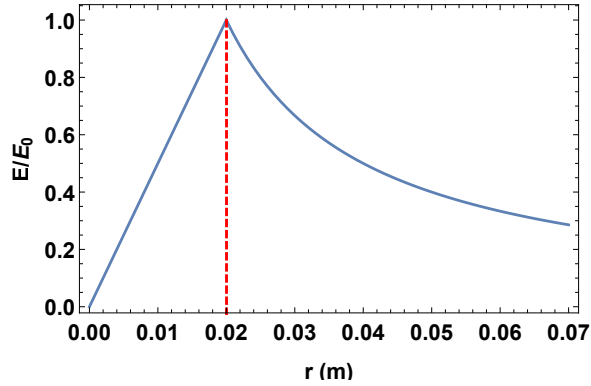}
\end{center}
\caption{Radial profile of wave electric field for step-like density shown in Fig.~\ref{fg_steplike}.}
\label{fg_surface}
\end{figure}

\begin{figure}[ht]
\begin{center}$
\begin{array}{ll}
(a)&(b)\\
\includegraphics[width=0.46\textwidth,angle=0]{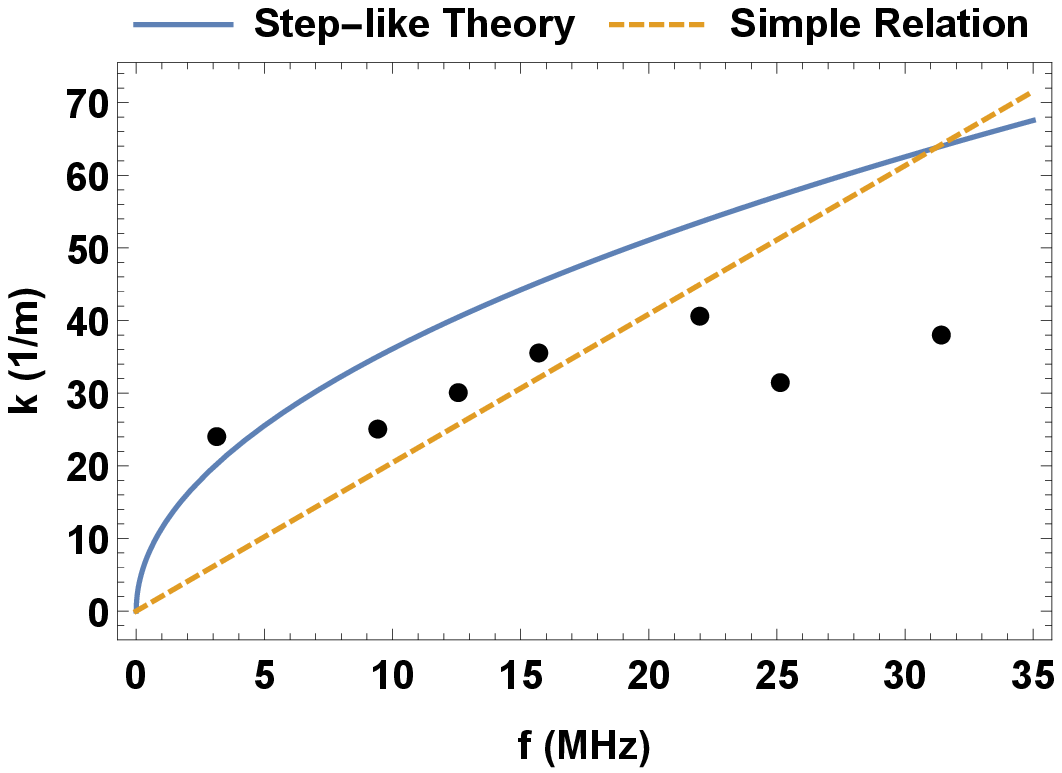}&\includegraphics[width=0.46\textwidth,angle=0]{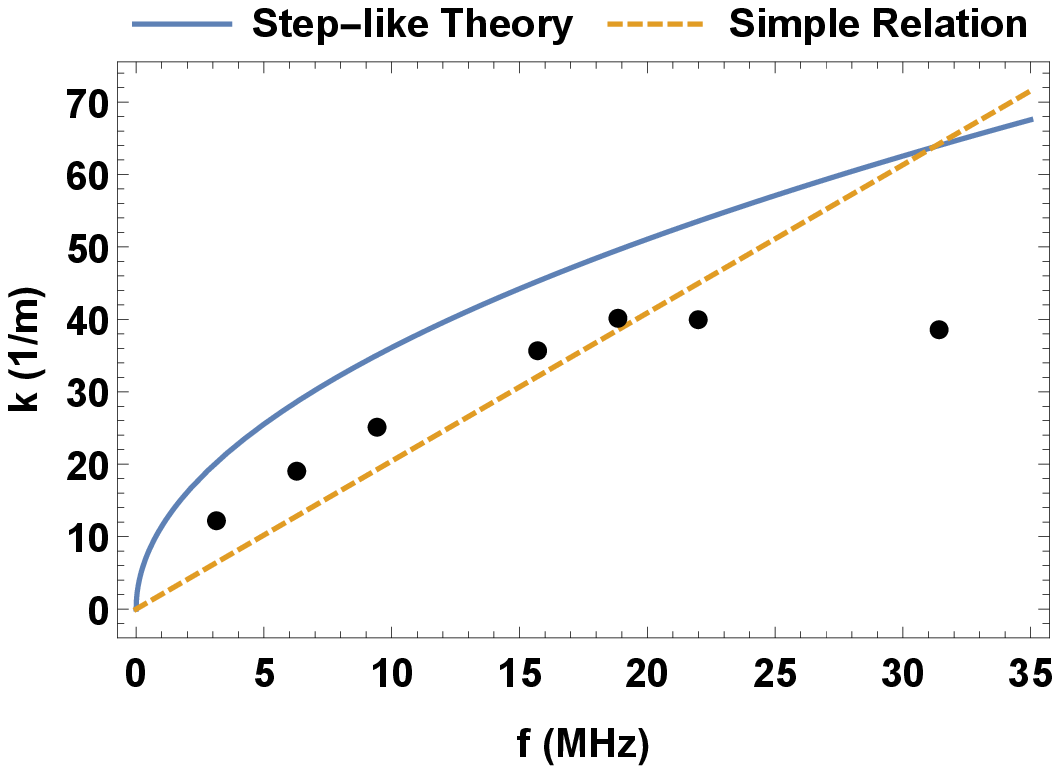}
\end{array}$
\end{center}
\caption{Dispersion relations inside (a, $r=0$~m) and outside (b, $r=0.04$~m) the blue-core column for $B_0=0.16$~T (step-like density profile).}
\label{fg_dispersion_sp}
\end{figure}

To reveal the underlying physics more clearly, we employ the step-like function theory developed by Breizman and Arefiev for radially localised helicon mode\cite{Breizman:2000aa}. The theory comprises two equations: 
\small
\begin{equation}
\frac{1}{r}\frac{\partial}{\partial r}\left[r\frac{\partial E}{\partial r}\right]-\frac{m^2}{r^2}E=-\frac{m}{k^2 r}\frac{\omega^2}{c^2}\frac{E\partial g/\partial r}{1+(m\partial g/\partial r)/k^2r\eta},
\label{eq16}
\end{equation}
\begin{equation}
\frac{1}{r}\frac{\partial}{\partial r}\left[\varepsilon r\frac{\partial}{\partial r}E_z\right]-\frac{m}{r}\left[\frac{\partial g}{\partial r}+\frac{\varepsilon m}{r}\right]E_z-k^2\eta E_z=0, 
\label{eq17}
\end{equation}
\normalsize
for helicon and TG modes, respectively, with $E=E_z\left[1+k^2 r\eta/(m\partial g/\partial r)\right]$. To focus on the radial density gradient, an artificial step-like density profile can be constructed:
\small
\begin{equation}
n_\alpha(r)=\left\{
\begin{array}{ll}
n_0~~~&\textrm{for}~~r<r_\ast,\\[5pt]
n_\ast~~~&\textrm{for}~~r>r_\ast.
\end{array}
\right.
\label{eq18}
\end{equation}
\normalsize
The symbols of $n_0$ and $r_0$ represent the maximum density on axis and edge radius of plasma column, respectively, while $n_\ast$ and $r_\ast$ label the magnitude and radius of density jump. Figure~\ref{fg_steplike} shows an illustration. Please note that this step-like density profile, although is unrealistic in experiment, can manifest the effect of radial density gradient and simplify the mathematical treatment. 
The resulted electric field has expression\cite{Breizman:2000aa}:
\small
\begin{equation}
E(r)=E_0\left\{
\begin{array}{ll}
(r/r_\ast)^{|m|}~~~&\textrm{for}~~r<r_\ast,\\[5pt]
(r/r_\ast)^{-|m|}~~~&\textrm{for}~~r>r_\ast,
\end{array}
\right.
\label{eq19}
\end{equation}
\normalsize
with $E_0$ a constant. For the $m=1$ mode considered here, its radial profile is displayed in Fig~\ref{fg_surface} (with $r_\ast=0.02$~m and $r_0=0.07$~m referring to the CSDX experiments\cite{Thakur:2015aa, Thakur:2014aa}). 
This resembles the computed results shown in Fig.~\ref{fg_efd_gsr} for the blue-core mode (high field cases) that wave electric field peaks off axis and decreases toward both axis and edge. The analytical dispersion relation is accordingly
\small
\begin{equation}
\omega=2\frac{m}{|m|}\frac{\omega_{ce}k^2 c^2}{\omega_{pe}^2(n_\ast)-\omega_{pe}^2(n_0)}. 
\label{eq20}
\end{equation}
\normalsize
To compare with numerical results, again we run the EMS code for various frequencies ($f=\pi\sim 10\pi$~MHz) and employ Fourier decomposition method to extract the dispersion relations, similar to the procedure done for Fig.~\ref{fg_dispersion_gs}. Here, the employed conditions include $n_\ast=0.3\times n_0$ and $r_\ast=0.3\times r_0$. As shown in Fig.~\ref{fg_dispersion_sp}, the dispersion curves from step-like theory (Eq.~(\ref{eq20})) and simple relation (Eq.~(\ref{eq15})) agree with the computed results (dots) qualitatively both inside and outside the blue-core column. Surprisingly, the simple relation based on slab geometry and single density value (on axis here) exhibits slightly better consistence, as observed in previous studies\cite{Chang:2012aa, Chang:2020aa}; moreover, as long as the blue-core column has been formed, we observe interference on the axial profiles of wave field for both Gaussian and step-like density configurations, which indicates the existence of multiple wave modes inside the blue-core column. Comparing Fig.~\ref{fg_dispersion_sp}(b) with Fig.~\ref{fg_dispersion_gs}(b), one could find the difference here that the wave number outside the blue-core column is not constant but varies with driving frequency. This may be caused by the density level outside the blue-core column which is not vanishing here but $0.3$ of the peak value on axis.
Next, to explore in detail the effects of density jump, i. e. magnitude and location, we introduce the ratios of $C_n=n_\ast/n_0$ and $C_r=r_\ast/r_0$, respectively, and the product of $C_b=C_n\times C_r$ to quantify the shrinking feature of blue-core plasma, as shown in Fig.~\ref{fg_steplike}. This product can be defined in phrase of ``blue-core constant" or more generally ``bright-core constant" to cover other gases as well. 
\begin{figure}[ht]
\begin{center}$
\begin{array}{ll}
(a)&(b)\\
\includegraphics[width=0.46\textwidth,angle=0]{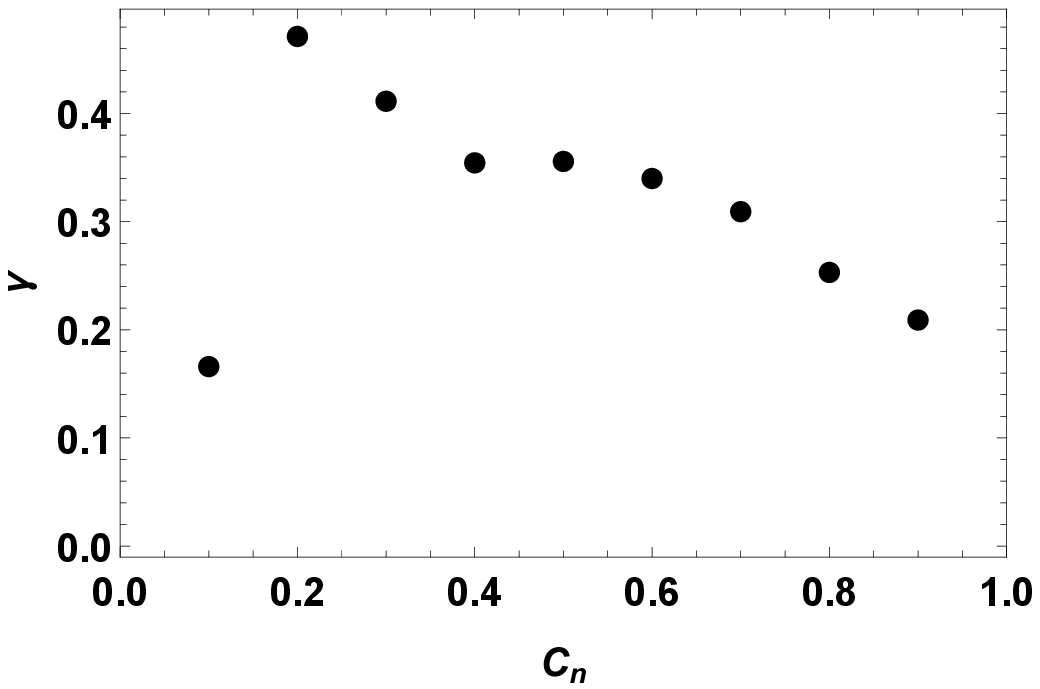}&\includegraphics[width=0.46\textwidth,angle=0]{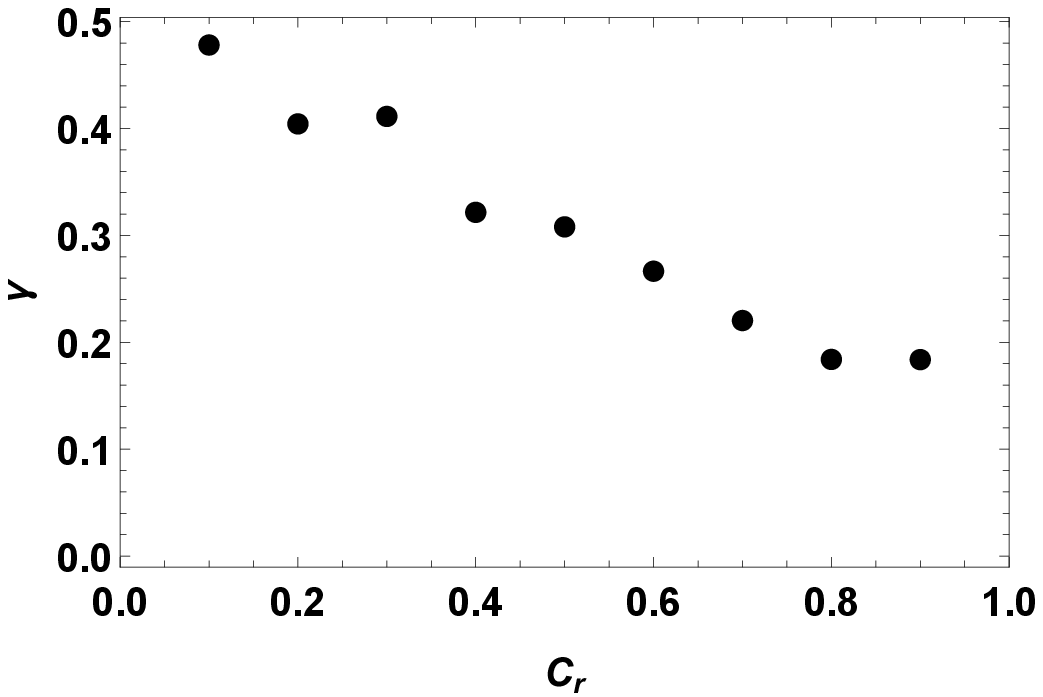}
\end{array}$
\end{center}
\caption{Dependence of $\gamma$ ($\gamma=P_\ast/P_0$) on: (a) $C_n=n_\ast/n_0$, (b) $C_r=r_\ast/r_0$.}
\label{fg_cb}
\end{figure}
We also setup the parameter of $\gamma=P_\ast/P_0$ in simulations to measure the ratio of power deposition inside the blue-core column ($P_\ast$) to that in total ($P_0$). The radius for $P_\ast$ is chosen to be $r=0.02$~m, same to the measured edge of blue-core column in experiments\cite{Thakur:2015aa, Thakur:2014aa}. Figure~\ref{fg_cb} shows the computed dependence of $\gamma$ on $C_n$ and $C_r$. We can see that $\gamma$ largely increases with reduced $C_n$ and $C_r$, which is expected because the plasma column is more shrunk for smaller $C_n$ and $C_r$, except for $C_n<0.2$ the ratio drops back. This could be attributed to the plasma density near edge which is too low to efficiently couple the power from antenna into core. This critical role of edge density has been also claimed by other studies\cite{Chang:2016aa, Chang:2018ab, Wang:2019aa, Isayama:2019aa, Isayama:2019ab}. 

\subsection{Equivalence to optical fiber}
\begin{figure}[ht]
\begin{center}
\includegraphics[width=0.8\textwidth,angle=0]{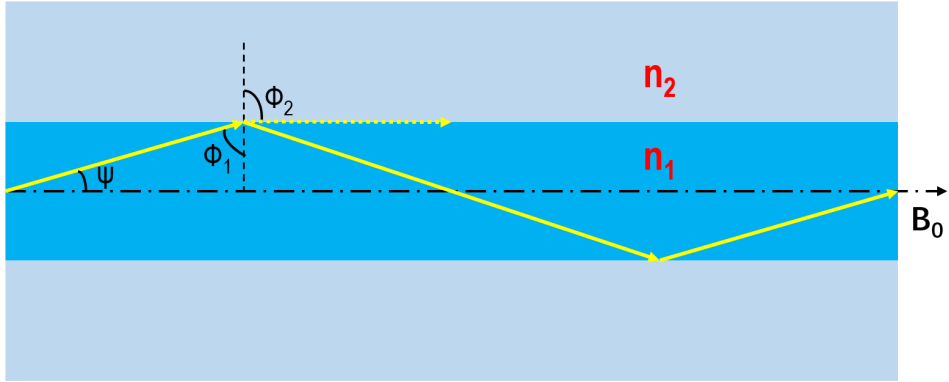}
\end{center}
\caption{Illustration of blue-core plasma column behaving as ``optical fiber" for electromagnetic communications.}
\label{fg_fiber}
\end{figure}

\begin{figure}[ht]
\begin{center}
\includegraphics[width=0.8\textwidth,angle=0]{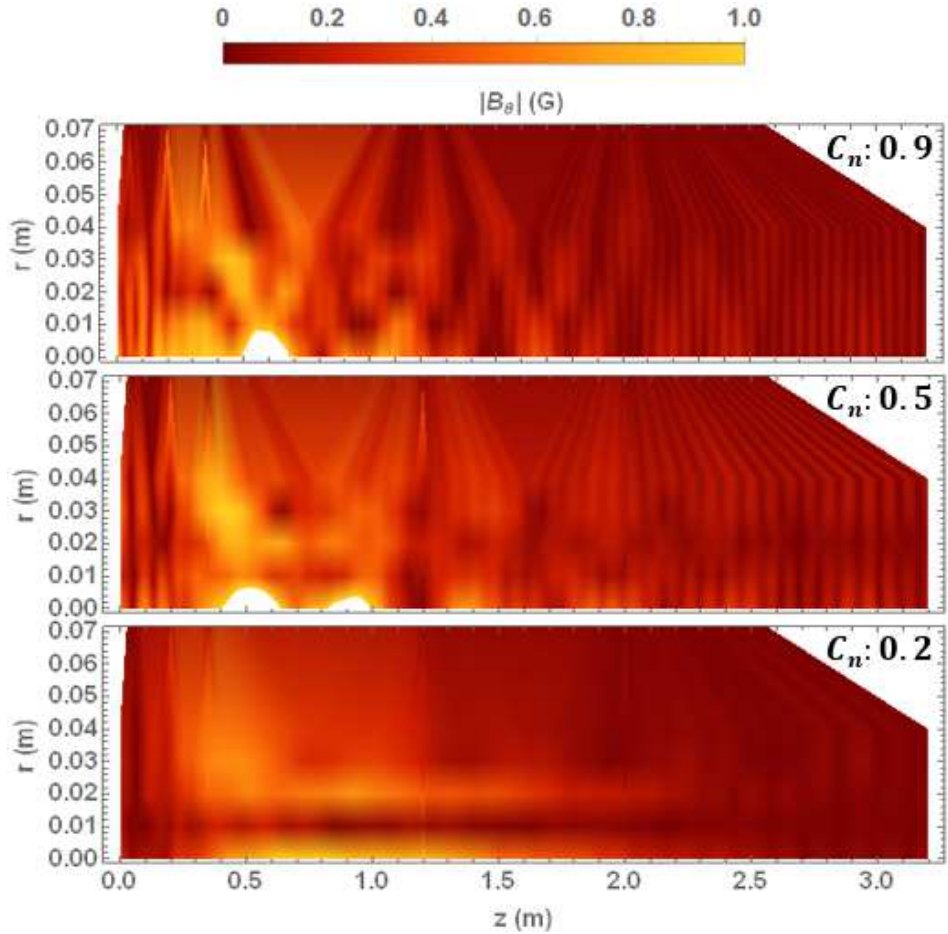}
\end{center}
\caption{2D Wave amplitude ($B_\theta$) for different values of $C_n$ ($C_r$ fixed to $0.3$).}
\label{fg_amplitude}
\end{figure}

Inspired by the finding above that wave propagation is confined mostly inside the blue-core column and the radial density gradient is very large (close to step-like), we propose that this blue-core helicon plasma could be used for electromagnetic communications, similar to the optical fiber for light communication\cite{Kao:1966aa, Maurer:1973aa}. To achieve this, the condition of total reflection has to be met, and Fig.~\ref{fg_fiber} gives an illustration. This illustration is consistent with previous findings that obliquely propagating waves are reflected by radial density gradient before they reach the edge of plasma\cite{Samm:2008aa, Caneses:2015aa}, following a zig-zap motion confined to the high density core. 
According to the law of refraction, namely $n_1 \sin \phi_1=n_2\sin\phi_2$ with $n$ the index of refraction and $\phi$ the angle to normal direction, we know that the threshold angle for total reflection ($\phi_2=\pi/2$) is $\phi_1=\textrm{arcsin}~(n_2/n_1)$. Referring to the definition of $n$ ($n=c/v_{ph}$ or $n=\lambda_0/\lambda$ with $c$ the speed of light, $v_{ph}$ the phase velocity, and $\lambda_0$ the wavelength in vacuum) and waves in uniform magnetised plasma\cite{Ginzburg:1970aa, Gurnett:2005aa}, we can write $n$ in form of 
\small
\begin{equation}
n^2=\frac{G\pm F}{2(\varepsilon \sin^2\psi+\eta \cos^2\psi)}
\label{eq21}
\end{equation}
\normalsize
with:
\small
\begin{equation}
G=(\varepsilon^2-g^2)\sin^2\psi+\varepsilon\eta(1+\cos^2\psi),
\label{eq22}
\end{equation}
\begin{equation}
F^2=\left[(\varepsilon^2-g^2)-\varepsilon\eta\right]^2\sin^4\psi+4 g^2\eta^2\cos^2\psi.
\label{eq23}
\end{equation}
\normalsize
Please note that here $\psi$ labels the angle of wave vector to the confining magnetic field, which lies in the same direction of blue-core edge, so that we have $\psi+\phi_1=\pi/2$ as shown in Fig.~\ref{fg_fiber}. For either parallel wave, i. e. 
\small
\begin{equation}
n^2=1-\sum_\alpha\frac{\omega_{p\alpha}^2}{\omega(\omega\pm\omega_{c\alpha})}\approx \sum_\alpha\frac{\omega_{p\alpha}^2}{\omega(\omega\pm\omega_{c\alpha})}, 
\label{eq24}
\end{equation}
\normalsize
or oblique wave such as whistler mode, i. e.
\small
\begin{equation}
n^2=\frac{\omega_{pe}^2}{\omega(\omega_{ce}\cos\psi-\omega)}, 
\label{eq24}
\end{equation}
\normalsize
we can all draw conclusion with high-density approximation ($\omega^2\ll\omega_{p\alpha}^2$ and $\omega_{c\alpha}^2\ll\omega_{p\alpha}^2$) that the index of refraction is proportional to the square root of plasma density, i. e. $n\propto \sqrt{n_{\alpha}}$, if other conditions (frequency and field strength) are fixed. Therefore, as long as the incident angle is bigger than the threshold value of $\phi_1=\textrm{arcsin}~(\sqrt{n_{\ast}}/\sqrt{n_{0}})$, the total reflection will occur and the blue-core plasma can indeed behave as an ``optical fiber" for electromagnetic communications. Figure~\ref{fg_amplitude} plots the two-dimensional wave amplitude ($B_\theta$) for different values of $C_n$ ($C_r$ fixed to $0.3$). We can see that with increased magnitude of density jump, i. e. decreased $C_n$ ($=n_\ast/n_0$), there forms a clear boundary layer around the location of density jump ($r=0.021$~m) which separates the cylinder radially into two regions: inner and outer. While the bright area near the boundary layer may be caused by wave reflections from antenna and core, the bright area near axis confirms that wave propagation is well confined inside the inner, as long as the density jump is sufficient, a feature same to optical fiber. 
Similarly, the blue-core plasma could also act as a waveguide for waves of other frequencies\cite{Southworth:1950aa}. Experimental verification of these ideas can be implemented on a well-defined blue-core helicon plasma with a tiny wave-exciting antenna located inside the central column and multiple receiving antennas placed outside. We leave it as a separate study and shall present in the future. 

\section{Conclusion}
The mechanism of blue-core phenomenon during helicon discharge has been an enigma for the research community. Different from existing studies which mainly employ optical camera or spectrometer to capture the formation procedure experimentally, this work devotes itself to revealing the detailed physics through computing the wave propagation and power deposition characteristics, referring to the recent experiments on CSDX\cite{Thakur:2015aa, Thakur:2014aa}. A well-benchmarked electromagnetic solver, based on Maxwell's equations and a cold-plasma dielectric tensor, is made use of. We found that: (i) the wave electric field peaks off-axis and near the radial location of particle transport barrier observed in experiment, an evidence of radial electrostatic confinement, during the blue-core formation; (ii) the wave magnetic field shows multiple radial modes inside the blue-core column, consistent with the experimental observation of coherent high $m$ modes through Bessel function; (iii) the axial profiles of wave field demonstrate that, once the blue-core mode has been established, waves can only propagate inside the central column with identical phase in the axial direction, agreeing with the early finding of light emission in phase axially\cite{Boswell:2021aa}; (iv) the two-dimensional distributions of wave energy and power deposition show off-axis (or hollow) features in radius, especially far from antenna in the diffusion region, and periodic structures in axial direction besides the blue-core boundary layer with either double (wave energy) or the same (power deposition) periodicity; (v) the analysis using step-like function theory provides consistent results, in terms of the radial profile of wave electric field and dispersion relation inside the blue-core column; (vi) the equivalence of blue-core helicon plasma column to optical fiber for electromagnetic communications possesses theoretical feasibility, as long as the incident angle is larger than the threshold value, e. g. $\phi_1=\textrm{arcsin}~(\sqrt{n_{\ast}}/\sqrt{n_{0}})$, and inspires novel applications of helicon plasma. Future research may be devoted to the experimental verification of this ``blue-core fiber" and the physics modelling of a more challenging topic: high-beta effects. 

\ack
We appreciate Dr. Guangye Chen for providing the EMS code and many instructions for its usage. This work was supported by the Chinese Academy of Sciences ``$100$" Talent Program (B) and the Science Foundation of Institute of Plasma Physics (DSJJ-$2020$-$07$).

\section*{Data Availability Statement}
The data that support the findings of this study are available from the authors upon reasonable request.

\section*{ORCID IDs}
\raggedright
Lei Chang: https://orcid.org/0000-0003-2400-1836

Juan F. Caneses: https://orcid.org/0000-0001-6123-2081

Saikat C. Thakur: https://orcid.org/0000-0002-8422-2705

Huai-Qing Zhang: https://orcid.org/0000-0002-2631-4740

\section*{References}
\bibliographystyle{unsrt}

\begin{thebibliography}{10}

\bibitem{Boswell:1970aa}
R.~W. Boswell.
\newblock Plasma production using a standing helicon wave.
\newblock {\em Physics Letters A}, 33(7):457--458, 1970.

\bibitem{Boswell:1997aa}
R.~W. Boswell and F.~F. Chen.
\newblock Helicons-the early years.
\newblock {\em IEEE Transactions on Plasma Science}, 25(6):1229--1244, 1997.

\bibitem{Chen:1997aa}
F.~F. Chen and R.~W. Boswell.
\newblock Helicons-the past decade.
\newblock {\em IEEE Transactions on Plasma Science}, 25(6):1245--1257, 1997.

\bibitem{Chen:2015aa}
F.~F. Chen.
\newblock Helicon discharges and sources: a review.
\newblock {\em Plasma Sources Science and Technology}, 24(1):014001, 2015.

\bibitem{Shinohara:2018aa}
S.~Shinohara.
\newblock Helicon high-density plasma sources: physics and applications.
\newblock {\em Advances in Physics: X}, 3(1):1420424, 2018.

\bibitem{Takahashi:2019aa}
K.~Takahashi.
\newblock Helicon-type radiofrequency plasma thrusters and magnetic plasma
  nozzles.
\newblock {\em Reviews of Modern Plasma Physics}, 3(1):3, 2019.

\bibitem{Guo:1999aa}
X.~M. Guo, J.~Scharer, Y.~Mouzouris, and L.~Louis.
\newblock Helicon experiments and simulations in nonuniform magnetic field
  configurations.
\newblock {\em Physics of Plasmas}, 6(8):3400, 1999.

\bibitem{Blackwell:2012aa}
B.~D. Blackwell, J.~F. Caneses, C.~M. Samuell, J.~Wach, J.~Howard, and C.~Corr.
\newblock Design and characterization of the magnetized plasma interaction
  experiment (magpie): a new source for plasma--material interaction studies.
\newblock {\em Plasma Sources Science and Technology}, 21(5):055033, 2012.

\bibitem{Thakur:2015aa}
S.~C. Thakur, C.~Brandt, L.~Cui, J.~J. Gosselin, and G.~R. Tynan.
\newblock Formation of the blue core in argon helicon plasma.
\newblock {\em {IEEE} Transactions on Plasma Science}, 43(8):2754--2759, 2015.

\bibitem{Zhang:2021aa}
T.~L. Zhang, R.~L. Cui, W.~Y. Zhu, Q.~Yuan, J.~T. Ouyang, K.~Y. Jiang, H.~B.
  Zhang, C.~W. Wang, and Q.~Chen.
\newblock Influence of neutral depletion on blue core in argon helicon plasma.
\newblock {\em Physics of Plasmas}, 28(7):073505, 2021.

\bibitem{Wang:2021aa}
C.~W. Wang, Y.~Liu, M.~Sun, T.~L. Zhang, Q.~Chen, and H.~B. Zhang.
\newblock Effect of inhomogeneous magnetic field on blue core in ar helicon
  plasma.
\newblock {\em Physics of Plasmas}, 28(12):123519, 2021.

\bibitem{Chang:2022aa}
L.~Chang, R.~Boswell, and G.~N. Luo.
\newblock First helicon plasma physics and applications workshop.
\newblock {\em Frontiers in Physics}, 9:808971, 2022.

\bibitem{Zhao:2017aa}
G.~Zhao, H.~H. Wang, X.~L. Si, J.~T. Ouyang, Q.~Chen, and C.~Tan.
\newblock The discharge characteristics in nitrogen helicon plasma.
\newblock {\em Physics of Plasmas}, 24(12):123507, 2017.

\bibitem{Huang:2020aa}
T.~Y. Huang, C.~G. Jin, Y.~W. Yu, J.~S. Hu, J.~H. Yang, F.~Ding, X.~H. Chen,
  P.~Y. Ji, J.~W. Qian, J.~J. Huang, B.~Yu, and X.~M. Wu.
\newblock Helicon-wave-excited helium plasma performance and wall-conditioning
  study on {EAST}.
\newblock {\em IEEE Transactions onPlasma Science}, 48(8):2878--2883, 2020.

\bibitem{Boswell:1984ab}
R.~W. Boswell.
\newblock Very efficient plasma generation by whistler waves near the lower
  hybrid frequency.
\newblock {\em Plasma Physics and Controlled Fusion}, 26(10):1147, 1984.

\bibitem{Boswell:2021aa}
R.~W. Boswell.
\newblock Helicon sources: Why they work!
\newblock In {\em First Helicon Plasma Physics and Applications Workshop},
  Hefei, China, 22-25 Sep. 2021.

\bibitem{Chen:2006aa}
G.~Chen, A.~V. Arefiev, R.~D. Bengtson, B.~N. Breizman, C.~A. Lee, and L.~L.
  Raja.
\newblock Resonant power absorption in helicon plasma sources.
\newblock {\em Physics of Plasmas}, 13(12):123507, 2006.

\bibitem{Thakur:2014aa}
S.~C. Thakur, C.~Brandt, L.~Cui, J.~J. Gosselin, A.~D. Light, and G.~R. Tynan.
\newblock Multi-instability plasma dynamics during the route to fully developed
  turbulence in a helicon plasma.
\newblock {\em Plasma Sources Science and Technology}, 23(4):044006, 2014.

\bibitem{Breizman:2000aa}
B.~N. Breizman and A.~V. Arefiev.
\newblock Radially localized helicon modes in nonuniform plasma.
\newblock {\em Physical Review Letters}, 84(17):3863--3866, 2000.

\bibitem{Corr:2007aa}
C.~S. Corr and R.~W. Boswell.
\newblock High-beta plasma effects in a low-pressure helicon plasma.
\newblock {\em Physics of Plasmas}, 14(12):122503, 2007.

\bibitem{Zhang:2008aa}
Y.~Zhang, W.~W. Heidbrink, H.~Boehmer, R.~McWilliams, G.~Chen, B.~N. Breizman,
  S.~Vincena, T.~Carter, D.~Leneman, W.~Gekelman, P.~Pribyl, and B.~Brugman.
\newblock {{Spectral gap of shear Alfv\'{e}n waves in a periodic array of
  magnetic mirrors}}.
\newblock {\em Physics of Plasmas}, 15(1):012103, 2008.

\bibitem{Lee:2011aa}
C.~A. Lee, G.~Chen, A.~V. Arefiev, R.~D. Bengtson, and B.~N. Breizman.
\newblock Measurements and modeling of radio frequency field structures in a
  helicon plasma.
\newblock {\em Physics of Plasmas}, 18(1):013501, 2011.

\bibitem{Chang:2012aa}
L.~Chang, M.~J. Hole, J.~F. Caneses, G.~Chen, B.~D. Blackwell, and C.~S. Corr.
\newblock Wave modeling in a cylindrical non-uniform helicon discharge.
\newblock {\em Physics of Plasmas}, 19(8):083511, 2012.

\bibitem{Chang:2013aa}
L.~Chang, B.~N. Breizman, and M.~J. Hole.
\newblock Gap eigenmode of radially localized helicon waves in a periodic
  structure.
\newblock {\em Plasma Physics and Controlled Fusion}, 55(2):025003, 2013.

\bibitem{Chang:2018aa}
L.~Chang.
\newblock Preliminary computation of the gap eigenmode of shear alfv{\'{e}}n
  waves on the {LAPD}.
\newblock {\em Chinese Physics B}, 27(12):125201, 2018.

\bibitem{Chang:2020aa}
L.~Chang, J.~Liu, X.~G. Yuan, X.~Yang, H.~S. Zhou, G.~N. Luo, X.~J. Zhang,
  Y.~K. Peng, J.~Dai, and G.~R. Hang.
\newblock Helicon plasma in a magnetic shuttle.
\newblock {\em {AIP} Advances}, 10(10):105114, 2020.

\bibitem{Ginzburg:1970aa}
V.~L. Ginzburg.
\newblock {\em The propagation of electromagnetic waves in plasmas}.
\newblock Pergamon Press, second edition, 1970.

\bibitem{Burin:2005aa}
M.~J. Burin, G.~R. Tynan, G.~Y. Antar, N.~A. Crocker, and C.~Holland.
\newblock On the transition to drift turbulence in a magnetized plasma column.
\newblock {\em Physics of Plasmas}, 12(5):052320, 2005.

\bibitem{Cui:2016aa}
L.~Cui, A.~Ashourvan, S.~C. Thakur, R.~Hong, P.~H. Diamond, and G.~R. Tynan.
\newblock Spontaneous profile self-organization in a simple realization of
  drift-wave turbulence.
\newblock {\em Physics of Plasmas}, 23(5):055704, 2016.

\bibitem{Lundin:2006aa}
R.~Lundin and A.~Guglielmi.
\newblock Ponderomotive forces in cosmos.
\newblock {\em Space Science Reviews}, 127(1-4):1--116, 2006.

\bibitem{Khazanov:2000aa}
G.~V. Khazanov, I.~K. Khabibrakhmanov, and E.~N. Krivorutsky.
\newblock Interaction between an alfv{\'{e}}n wave and a particle undergoing
  acceleration along a magnetic field.
\newblock {\em Physics of Plasmas}, 7(1):1--4, 2000.

\bibitem{Khazanov:2013aa}
G.~V. Khazanov and E.~N. Krivorutsky.
\newblock Ponderomotive force in the presence of electric fields.
\newblock {\em Physics of Plasmas}, 20(2):022903, 2013.

\bibitem{Chang:2014aa}
L.~Chang.
\newblock {\em The impact of magnetic geometry on wave modes in cylindrical
  plasmas}.
\newblock PhD thesis, Australian National University, Canberra, 2014.

\bibitem{Caneses:2016aa}
J.~F. Caneses and B.~D. Blackwell.
\newblock Collisional damping of helicon waves in a high density hydrogen
  linear plasma device.
\newblock {\em Plasma Sources Science and Technology}, 25(5):055027, 2016.

\bibitem{Shamrai:1996aa}
K.~P. Shamrai and V.~B. Taranov.
\newblock Volume and surface rf power absorption in a helicon plasma source.
\newblock {\em Plasma Sources Science and Technology}, 5(3):474--491, 1996.

\bibitem{Caneses:2017aa}
J.~F. Caneses, B.~D. Blackwell, and P.~Piotrowicz.
\newblock Helicon antenna radiation patterns in a high-density hydrogen linear
  plasma device.
\newblock {\em Physics of Plasmas}, 24(11):113513, 2017.

\bibitem{Chen:1996ab}
F.~F. Chen.
\newblock Physics of helicon discharges.
\newblock {\em Physics of Plasmas}, 3(5):1783--1793, 1996.

\bibitem{Kao:1966aa}
K.~C. Kao and G.~A. Hockham.
\newblock Dielectric-fibre surface waveguides for optical frequencies.
\newblock {\em Proceedings of the Institution of Electrical Engineers},
  113(7):1151--1158, 1966.

\bibitem{Maurer:1973aa}
R.~D. Maurer.
\newblock Glass fibers for optical communications.
\newblock {\em Proceedings of the {IEEE}}, 61(4):452--462, 1973.

\bibitem{Chang:2016aa}
L.~Chang, Q.~C. Li, H.~J. Zhang, Y.~H. Li, Y.~Wu, B.~L. Zhang, and Z.~Zhuang.
\newblock Effect of radial density configuration on wave field and energy flow
  in axially uniform helicon plasma.
\newblock {\em Plasma Science and Technology}, 18(8):848--854, 2016.

\bibitem{Chang:2018ab}
L.~Chang, X.~Y. Hu, L.~Gao, W.~Chen, X.~M. Wu, X.~F. Sun, N.~Hu, and C.~X.
  Huang.
\newblock Coupling of {RF} antennas to large volume helicon plasma.
\newblock {\em AIP Advances}, 8(4):045016, 2018.

\bibitem{Wang:2019aa}
R.~L. Wang, L.~Chang, X.~Y. Hu, L.~L. Ping, N.~Hu, X.~M. Wu, J.~Y. Yao, X.~F.
  Sun, and T.~P. Zhang.
\newblock The role of second-order radial density gradient for helicon power
  absorption.
\newblock {\em Contributions to Plasma Physics}, 59(9):e201900032, 2019.

\bibitem{Isayama:2019aa}
S.~Isayama, S.~Shinohara, T.~Hada, and S.~H. Chen.
\newblock Underlying competition mechanisms in the dynamic profile formation of
  high-density helicon plasma.
\newblock {\em Physics of Plasmas}, 26(2):023517, 2019.

\bibitem{Isayama:2019ab}
S.~Isayama, S.~Shinohara, T.~Hada, and S.~H. Chen.
\newblock Spatio-temporal behavior of density jumps and the effect of neutral
  depletion in high-density helicon plasma.
\newblock {\em Physics of Plasmas}, 26(5):053504, 2019.

\bibitem{Samm:2008aa}
U.~Samm.
\newblock Plasma-wall interaction in magnetically confined fusion plasmas.
\newblock {\em Fusion Science and Technology}, 53(2T):223--228, 2008.

\bibitem{Caneses:2015aa}
J.~F. Caneses.
\newblock {\em Helicon wave propagation and plasma equilibrium in high-density
  hydrogen plasma in converging magnetic fields}.
\newblock PhD thesis, Australian National University, 2015.
\newblock pp. 175.

\bibitem{Gurnett:2005aa}
D.~A. Gurnett and A.~Bhattacharjee.
\newblock {\em Introduction to {P}lasma {P}hysics with {S}pace and {L}aboratory
  {A}pplications}.
\newblock Cambridge University Press, 2005.

\bibitem{Southworth:1950aa}
G.~C. Southworth.
\newblock Principles and applications of waveguide transmission.
\newblock {\em Bell System Technical Journal}, 29(3):295--342, 1950.

\end{thebibliography}

\end{document}